
\newcount\fcount \fcount=0
\def\ref#1{\global\advance\fcount by 1
\global\xdef#1{\relax\the\fcount}}

\def\la{\lower.5ex\hbox{$\; \buildrel < \over \sim \;$}}
\def\ga{\lower.5ex\hbox{$\; \buildrel > \over \sim \;$}}

\magnification=\magstep1

\raggedbottom
\footline={\hss\tenrm\folio\hss}
\tolerance = 30000
\def\etal{{\it et al.}}

\baselineskip=5.0mm plus 0.1mm minus 0.1mm
\centerline{\bf COSMIC RAY IONIZATION OF THE INTERSTELLAR MEDIUM}
\vskip 1.5cm
\centerline {\bf Biman B. Nath and Peter L. Biermann}
\vskip 0.5cm
\centerline{Max Planck Institut f\"ur Radioastronomie,
Auf dem H\"ugel 69, D-53121 Bonn, Germany}
\vskip 0.5cm
\centerline{Accepted Nov. 15 1993;
Received August 24 1993}
\vskip 0.2cm
\centerline{To appear in {\it MNRAS}}
\vskip 2.0cm
\centerline {\bf Abstract}
We consider a cosmic ray spectrum that is a power law in momentum
down to a cutoff and derive a lower cutoff corresponding to
$E_{kin} \sim (30-60)$ MeV from the observed ionization rates in
nearby diffuse clouds. While the real spectra of cosmic rays may not
be so simple, we argue that one expects a substantial change in the
spectra at such energies and that, at first approximation, a power
law spectra with a lower cutoff is appropriate. Such a description of
the cosmic rays in the interstellar medium is not only theoretically
more attractive than the spectra used in the literature, but is also
supported by recent observations.
 \vskip 2.0cm

\noindent {\bf Introduction} \medskip
The importance of cosmic rays in heating the intersteller medium (ISM)
has been recognized since long. Hayakawa, Nishimura and Takayanagi
(1961) calculated the flux of low-energy cosmic rays required to
account for a kinetic temperature of
$125$ K, employing a power law in energy that is observed at
high energies and extrapolating it down to $\sim 10$ MeV. Spitzer
 and Tomasko (1968) reconsidered the problem in the
light of observations of lesser temperatures ($\sim 60$ K) and
calculated the ionization rate for the demodulated spectrum of cosmic
rays to be $6.8 \times 10^{-18}$ sec$^{-1}$. They
determined a heating rate of $\sim 5.7$ eV per ionization event.
Field {\it et al.} (1969) assumed a cosmic ray
energy density that corresponds to an ionization rate of $4 \times
10^{-16}$ sec$^{-1}$.

However, on one hand, recent observations of abundances in the ISM
point to an ionization rate much different from the numbers above.
Black {\it et al.} (1990) and van Dishoeck and Black (1991) recently
inferred an ionization rate of $(3-7) \times 10^{-17}$ sec$^{-1}$
from observations of nearby diffuse clouds. The dominant contributions
to the ionization,
especially for the abundances of species such as H$_3^+$,
are expected to come from low energy cosmic rays.
On the other hand, theoretical advances in the field of cosmic
ray acceleration have shown that the demodulated cosmic ray
spectrum, that pervades the ISM, is most probably a power law in momentum
($p$) (Krymsky 1977; Axford {\it et al.} 1977; Bell 1978; Blandford
and Ostriker 1978). This is very different from the spectrum
used by Spitzer and Tomasko (1968) or Field {\it et al.} (1969).
We assume here that the cosmic ray spectrum is a power law in momentum
and that any substantial change at the lower end can be approximated
as a lower cutoff. Then, the
ionization and heating will depend on (a) the spectral
index, (b) the amplitude and (c) the lower cutoff. Furthermore, if
one uses the spectral index and the amplitude as one observes
for the high energy tail, then the only parameter that remains
to be determined is the lower cutoff.
The lower cutoff for different ions will be different, depending on
the energy losses as they traverse the ISM.

The description of the low energy cosmic ray spectrum in terms of a
single lower cutoff is ad hoc but is the simplest. In reality, due to
various energy loss processes, the spectrum will deviate from the power law
at some energy scale. However we still lack any detail knowledge of the
spectrum at energies below a few hundred MeV. We therefore assume a lower
cutoff to describe the characteristic energy scale at which the spectrum
changes from the power law in momentum.

In this paper we propose to describe the ionization and
the heating of the ISM with a simple but theoretically
plausible cosmic ray spectrum. We infer the lower cutoff
by calculating the ionization rate and comparing with the
observations. The lower cutoff that we infer could be important in
understanding and predicting possible heating of gas in
other galaxies, and the intergalactic gas in the vicinity of our
Galaxy (Nath and Biermann 1993).

\bigskip \noindent {\bf Ionization by energetic particles} \medskip

\centerline {\it (a) Ionization cross-section}

The cross-section for ionization
($\sigma$) of
hydrogen atom in the ground state by an energetic particle with
velocity $v=c \beta$ and with atomic number $Z$ is given by
(Bethe 1933)
$$
\sigma={1.23 \times 10^{-20} Z^2 \over \beta^2} \Bigl (
6.2+log_{10} {\beta ^2 \over 1- \beta ^2} - 0.43 \beta^2 \Bigr )
\>{\rm cm}^2 \> . \eqno(1)
$$
The effect of secondary electrons that are released by the initial
ionization events has been dealt with by Dalgarno and Griffing (1958)
and Spitzer and Scott (1969). The first generation electrons give
rise to additional ionization and more electrons. For nearly neutral
gas or whose fractional ionization ($f$) is small, the total number of
secondary electrons produced in ${ 5 \over 3}$ times the number of
electrons produced by `first generation' ionization. This number
is a function of $f$ and drops down to $1.12$ for $f \sim 0.3$.
However, since $f$ is small in diffuse clouds, we will use
the value ${5 \over 3}$.

The ionization rate of H atoms due to the cosmic rays then could
be written as
$$
\zeta= {5 \over 3} \int 4 \pi n(p) \sigma dp, \eqno(2)
$$
where $n(p)dp$ is the intensity of cosmic rays per unit area
per second per unit of solid angle with momentum within the
range $dp$. This is essentially Spitzer and Tomasko's (1968)
eq. $(4)$, however, with a differential spectrum in momentum
and not in energy as they used.
\medskip
\centerline{\it (b) Cosmic rays}
\medskip
As mentioned in the Introduction, theoretical
attempts to explain the acceleration of cosmic rays in shocks
have shown that the resultant spectrum is one in momentum, $p$
(Krymsky 1977; Axford \etal 1977; Bell 1978; Blandford and
Ostriker 1978). Biermann a, b), Biermann and Cassinelli (1993),
Biermann and Strom (1993), Stanev {\it et al.} (1993), Rachen and
Biermann (1993), and Rachen {\it et al.} (1993)
 has discussed the theoretical spectrum and compared
it with airshower data. They calculate an injection spectrum of
$p^{-2.42 \pm 0.04}$ from the idea that cosmic rays up to about
10 TeV (particle energy for protons) are dominated by supernova
explosions in the ISM and beyond that by supernova explosions
into stellar winds (till EeV) and radio galaxies (beyond EeV).
The observed $p^{-2.75 \pm 0.04}$ spectrum below 10 TeV
is explained by arguing that the leakage time
due to diffusion in the ISM, which has a Kolmogorov spectrum of
turbulence, changes the index.

It is possible that there are other sources of low
energy cosmic rays in the ISM. However, the available data seems to be
adequately explained by the supernova explosions (see the references
above, especially, Biermann and Strom 1993). In the present work, we adopt a
minimalist approach and assume the supernova explosions to be the underlying
mechanism for low energy cosmic rays, which suggest a spectrum with a power law
in momentum.

The observed amplitudes of the cosmic rays of different
elements at high energy have been reported as (Weibel 1992),
$$\eqalignno{
\phi _H&=(9.81 \pm 0.32) \times 10^{-2} \Bigl ({E \over TeV} \Bigr )
^{-2.74 \pm 0.02} &\cr
\phi _{He}&=(6.03 \pm 0.19) \times 10^{-2} \Bigl ({E \over TeV} \Bigr )
^{-2.61 \pm 0.03} &\cr
\phi _{C}&=(1.09 \pm 0.15) \times 10^{-2} \Bigl ({E \over TeV} \Bigr )
^{-2.69 \pm 0.03} &\cr
\phi _{O}&=(1.87 \pm 0.08) \times 10^{-2} \Bigl ({E \over TeV} \Bigr )
^{-2.64 \pm 0.04} &\cr
\phi _{Fe}&=(1.82 \pm 0.016) \times 10^{-2} \Bigl ({E \over TeV} \Bigr )
^{-2.63 \pm 0.05} &(3)\cr}
$$
where the unit of $\phi$ is (m$^2$ s sr TeV/nucleus)$^{-1}$. These spectral
data are consistent with the possibility that at TeV energies He-Fe spectra
are all dominated by wind supernovae, with spectra like $E^{-8/3}$,
(Biermann 1993a). The
subscripts H, He, C, O and Fe refer to the ions of hydrogen, helium,
carbon, oxygen and iron respectively.
\medskip
\centerline{ \it (c) Lower cutoff}
\medskip

As the cosmic ray particles travel through the
IGM to us (i) the spectrum index is changed as mentioned above and (ii)
the low energy particles are deleted from the spectra due to various
losses. At energies below $\sim 100$ MeV, the important mechanisms
for energy loss are due to ionization and spallation of the nuclei.
The total fragmentation cross-section of a nucleus (atomic weight
$A_T > 1$) by a beam of protons for can be approximated as
(Berezinsky {\it et al.} 1990),
$$
\sigma_{sp}= 2 \times 10^{-25} (A_T ^{3/4} -0.7)^2 \>
{\rm cm}^2. \eqno(6)
$$
We compare the energy losses of He nuclei due to ionization and spallation
in fig. 1, in the units of per second per $n_{HI, ISM}$, the hydrogen atom
density in the ISM (for ionization),
and per $n_{p, CR}$, cosmic ray proton density (for spallation).
In reality, of course, $n_{HI, ISM} \gg
n_{p, cr}$.
It is clear that for $E_{kin} < 100$ MeV, ionization losses
outweigh the losses due to spallation.

Since the energy loss depends on the atomic number, we expect the
lower cutoffs of various ions to be different. We can estimate
the dependence of the cutoff on the atomic number $Z$ by considering
the underlying physical mechanism of diffusion.
 Consider an ISM
 with a spatial spectrum of turbulence given by $I(k)$. Here, $I(k)k$ is
 the
 energy density of the turbulence and $k$ is the wavenumber.
 Assuming that the most effective scattering takes place when the
turbulence has lengthscale comparable to $r_g$, the radius
 of gyration of the particle in a magnetic field of strengh $B$, one
 readily finds that the mean free path $\lambda$ is given by (Drury 1983),
 $$
 \lambda = r_g {B^2 / 8 \pi \over I(k) k}.\eqno(4)
 $$
 For a Kolmogorov spectrum ($I(k) \propto k^{-5/3}$), the mean
 free path $\lambda \propto r_g^{1/3}$. Since the radius
of gyration $r_g \propto p/Z \propto {(E_{kin} A)^{1/2} \over Z}$
for the non-relativistic particles (where $A \sim 2Z$ is the atomic weight).
The diffusion coefficient
of the particle) is, therefore,
$
\kappa \propto Z^{-1/3} A^{1/3} E_{kin}^{2/3}
$
Assuming that the particles come from the same distance, the diffusion
time
$
t \propto {1 \over \kappa}  \propto Z^{1/3} A^{-1/3} E_{kin}^{-2/3}.
$
The lower cutoff is proportional to the total energy
loss ($t \times {dE \over dt}$). Since energy loss due to
ionization (neglecting the slow logarithmic term)
${dE \over dt} \propto {Z^2 \over v}$, we can write
$$\eqalignno{
E_{kin, \> cutoff} &\propto Z^{7/3} A^{5/6} E_{kin}^{-7/6},
\; {\rm i.e.,} &\cr
E_{kin, \> cutoff} &\propto Z^{14/13} A^{5/13} \propto Z^{19/13}
\; .&(7)\cr}
$$

With this dependence of the lower cutoffs of different ions,
we can express the ionization of the cosmic rays in terms of
the lower cutoff of a single ion, say, proton.
In the regimes where spallation dominates over ionization energy
losses, the cutoff $E \propto Z^{29/14}$
($dE/dt$ in this case scales differently).

Note that, in reality the spectra will not have a sharp low
energy cutoff, mainly because the observed particles come from
various sources at different distances. However, here we attempt
to approximate any possible change or turnover in the spectra with
a single parameter, {\it i.e.}, a lower cutoff.

\medskip \centerline{\it (d) Rate of ionization} \medskip

Using eq. $(2)$ to calculate the
ionization rate of cosmic rays with the spectra as in $(3)$
(in momentum space) and with the lower cutoffs obeying the
proportionality rule of $(8)$, we show the results
in fig. 2. The observations of Black (1990) are plotted
along with the timescale for ionization ($1/ \zeta$) from
the eq. $(2)$ as a function of the lower cutoff in kinetic
energy of cosmic ray protons. The figure shows that a cutoff
in the range of $(30-60)$ MeV is implied by the observations.

\medskip \centerline{\it (e) Heating by the cosmic rays} \medskip

Ionization produces energetic secondary electrons in the ISM.
These electrons, in a weakly ionized gas, lose their energy
mainly by excitation and further ionizations until the energy
falls below $10.2$ eV. The average energy left to heat the gas
depends on these low energy electrons and their spectrum.
Spitzer and Tomasko (1968) calculated the average energy
from the proton spectrum they used, {\it viz.},
$$
j(E_{kin}) ({\rm particles} \> {\rm cm}^{-2} \> {\rm s}^{-1} \>
{\rm sr}^{-1} \> {\rm GeV/nucleon}^{-1})=
{0.9 \over (0.85 +E_{GeV})^{2.6}} {1 \over (1+0.01/E_{GeV})}
.\eqno(7)
$$
This spectrum was folded with the spectrum one gets for
a single proton (of momentum $p$), $n_e(E) dE \propto
E^{-2} p^{-1} dE$.
Since the spectrum that we propose to use has a lower cutoff,
at $\sim 30$ MeV as shown in the previous section, the spectrum
of the first generation electrons is not very clear. However,
it is known that the mean energy of the secondary electrons
is a very slow function of the incident proton energy and
can be taken as $36$ eV (Dalgarno and Griffing 1958).
If we, for simplicity, assume that all the secondary electrons
have this energy, then it is easy to calculate the average
energy following the prescription by Spitzer and Tomasko.
We find that a heat source of $6.3$ eV is associated with
each primary ionizing event. For comparison, Spitzer and
Tomasko, whose spectrum had no lower cutoff, found a value
of $5.7$ eV, and Field {\it et al.}, using a different
spectrum, $8.5$ eV.

Comparing with the total energy $36$ eV lost per free electron
produced, the above average energy ($6.3$ eV) implies
an efficiency of $\epsilon =0.175$ for heating. However
one must note that in the case of high fractional ionization
($n_e/n_H > 0.01$), the secondary electrons will lose
energy more efficiently due to interactions with abundant free
electrons.

\bigskip \noindent {\bf Discussion} \medskip

It is worth noting that the lower cutoff derived above
is also suggested by creation of light elements from spallation
by cosmic ray particles. Gilmore {et al.} (1992) noted that
the production ratio $N_{ 10_B + 11_B}/ N_{9_{Be}}$  increases rapidly
for particle energies below $\sim 30$ MeV as the cross-section
is dominated by resonances and so that the particle spectrum
ought to have a cutoff about $\sim 30 $ MeV..

The model of the heating and ionization of the ISM sketched
above is probably too simplistic. But we believe that such
a picture is closer to reality than the spectra used
by previous workers. Seo {\it et al.} (1991) analyzed the
data obtained during a solar minimum and concluded that
the spectra of cosmic ray ions are almost certainly power
laws in rigidity outside the solar system, at least down to
$\sim 700$ MeV. They
found that below this energy (for protons), the flux of
particles after solar modulation is lesser than what a single
power low would predict. They suggested that this may
be the hint of a break in the power law in below $ \sim 200$ MeV.
We have tried to describe the spectrum in terms of a minimum
number of parameters, that is, a single power law with a lower
cutoff. While the result of Seo {\it et al.} is highly
suggestive of such a picture, it is difficult to confirm it
without working out the complete mechanism of solar modulation
and comparing with their data.

It is possible that low energy cosmic rays are accelerated inside
the molecular clouds. Dogiel and Sharov (1990) showed that the
energy density inside the clouds could be $10-100$ times higher
than what is observed on the earth and this could imply a larger
low energy cutoff for the cosmic rays {\it inside} these clouds
for the same ionization rate.
However, the uncertainties involved motivated us to use the observed
fluxes only and derive a low energy cutoff for the ISM. One must
note that such a description in terms of a single value of the low energy
cutoff everywhere in the Galaxy is probably unrealistic, but is the
simplest one.

\bigskip\noindent{\bf Conclusion} \medskip

We have used the observed ionization rates in the ISM to
derive a low cutoff in the cosmic ray spectrum which is
a power law in momentum. We have derived the scaling
relations for the low cutoff for the ions as a function
of $Z$ from the consideration of energy loss in traversing
the ISM. We find that a cutoff in the kinetic energy of protons
in the range of $30-60$ MeV tallies with the observations.
We have noted that such a cutoff is also supported by
arguments that spallations by cosmic rays should not
produce too much light elements than are observed.

\bigskip\bigskip\bigskip\centerline{\bf Acknowledgments}\medskip
PLB wishes to thank Dr. J. R. Jokipii for many
discussions on this topic; this project was started with Dr. J. R.
Jokipii while PLB
was on a five months sabbatical at the University of Arizona,
Tucson in 1991. PLB also wishes to thank Dr. J. H. Black for
critical discussions on the CR-induced ionization rate in diffuse
interstellar clouds. BBN thanks the Max Planck Society for a fellowship.
High energy astrophysics with
PLB is supported by the DFG (Bi 191/9), the BMFT (DARA FKZ
50 OR 9202) and a NATO travel grant (CRG 910072).

\vskip 2.0cm
\centerline{\bf References}\bigskip
\obeylines{
1. Axford, W. I. {\it et al.} 1977, Proc. 15th ICRC (Plovdiv), vol. 11, p. 132.
2. Bell, A. R. 1978, MNRAS, 182, 443.
3. Berezinsky. V. S. {\it et al.} 1990, Astrophysics of Cosmic
Rays, (North Holland, Amsterdam), p. 59.
4. Bethe, H. 1933, Hdb. d. Phys. (Berlin: J. Springer),
24, Pt. 1, 491.
5. Biermann, P. L., Strittmatter 1987, ApJ, 322, 643.
6. Biermann, P. L. 1993a  A\&A, 271, 649.
7. Biermann, P. L. 1993b, in ``High Energy Astrophysics'', Ed. J.
Matthews,(World Scientific, Singapore), in press.
8. Biermann, P. L., J. P. Cassinelli 1993, A\&A, 277, 691.
9. Biermann, P. L., R. G. Strom 1993, A\&A, 275, 659.
10. Black, J. H. {\it et al.} 1990. ApJ, 358, 459.
11. Blandford, R. D., Ostriker, J. P 1978, ApJ, 227, L49.
12. Dalgarno, A., Griffing, G. W. 1958, Proc. Roy. Soc. London, A, 248, 415.
13. Drury, L.O'C 1983. Rep. Prog. in Phys., 46, 973.
14. Field, G. B. {\it et al.} 1969, ApJ, 155, L149.
15. Gilmore, G. {\it et al.} 1992. Nature, 357, 379.
16. Hayakawa, S., Nishimura, S., Takayanagi, K. 1961, PASJ, 13, 184.
17. Krymsky, G. F. 1977, Sov. Phys. Dokl., 22, 327.
18. Nath, B. B., Biermann, P. L. 1993, MNRAS, 265, 241.
19. Rachen, J. P., Biermann, P. L. 1993, A\&A, 272, 161.
20. Rachen, J. P. {\it et al.} 1993, A\&A, 273, 377.
21. Seo, E. S. {\it et al.} 1991, ApJ, 378, 763.
22. Spitzer, L., Tomasko, M. G. 1968. ApJ, 152, 971.
23. Spitzer, L., Scott, E. H. 1969, ApJ, 158, 161.
24. Stanev, T. {\it et al.} 1993, A\&A, 274, 902.
25. Van Dishoeck, E., Black, J. H. 1991, ApJ, 369, L9.
26. Wiebel, B. 1992, Unpublished HEGRA-note, Diploma-Thesis in Physics,
University of Wuppertal, Germany.
}

\vskip 2.5cm
\centerline{\bf Figure Captions}
Figure 1: The rate of loss of energy for an ion (Helium is taken
as an example here, since it is the lightest ion that loses energy by
spallation) is plotted against the kinetic energy of the
ion. For spallation, the rate (per second) corresponds to per
cosmic ray proton, and for ionization, per neutral hydrogen atom in the ISM.
The curves show that ionization losses dominate over those from
spallation for the energy range we consider here for the lower
cutoff.
\medskip

Figure 2: Ionization timescale (second) is plotted as a function
of the lower cutoff of the proton kinetic energy in the cosmic
ray spectrum. The horizontal lines correspond to the observed
limits on the ionization rate $(3-7) \times 10^{-17}$ s$^{-1}$.

\vfill\eject
\end